# International institutions and power politics in the context of Chinese "Belt and Road Initiative"

Author- Mandeep Singh Rai
Affiliation-GMAP 22,
The Fletchers School of Law and Diplomacy, Tufts University.
Corresponding author email address-
Mandeep.Rai@tufts.edu,msrai1980@gmail.com

1. **<u>Introduction.</u>** The subject of international institutions and power politics continues to occupy a central position in the field of International Relations and to the world politics. It revolves around key questions on how rising states, regional powers and small states leverage international institutions for achieving social, political, economic gains for themselves. Taking into account one of the rising powers China and the role of international institutions in the contemporary international politics, this paper aims to demonstrate, how in pursuit of power politics, various states (Small, Regional and Great powers) utilize international institutions by making them adapt to the new power realities critical to world politics.

The first section covers the basic definitions of what is meant by international institutions & power politics and aspect of collective mobilization. In the second section, I bring out four strategies: power bargaining, strategic co-optation, rhetorical coercion and principled persuasion which the emerging states like China employ to challenge international institutions status quo and make them adapt to the present norms. To respond to unfavourable shifts, states rely on variety of processes such as soft balancing, information, wedging, networks and employ them through international institutions to pursue power more effectively. The importance of leadership and symbolic capital as sources of power to mould influence of international institution's behaviour is discussed by highlighting corresponding examples in the context to China.

Each of the strategies and processes are explored and related to the ways in which Beijing has impelmented them within the gambit of regional organizations like SCO (Shanghai Cooperation Organization), international programs in particular BRI (Belt and Road Initiative) & international institutions like UN (United nations) to improve its status as an emerging hegemon. In conclusion, I predict the way forward for power politics by use of international institutions in relation to Chinese growing influence.

2. **International institutions, Power Politics and Collective Mobilization**. I adopt the following definitions of 'International institutions' and 'Power politics' given by Anders and Wivel. International institutions are "*associational clusters among states with some bureaucratic structures that create and manage self-imposed and other-imposed constraints on state policies and behaviors.*"[1] Power politics "*is the contestation among individual states using their particular resources and bargaining strengths to influence the structure of relations and the conduct of other actors."*[2] In the context of institutions, power politics involves efforts by states to influence the formulation, application, and enforcement of the rules and regulations of a given institution as well as the control of bureaucratic positions and allocation of resources within it.[3] Ripsman provides a neoclassical realist argument that international institutions are useful tools for legitimizing foreign policy decisions, overcoming domestic opposition, easing the demands on domestic resource mobilization through burden sharing and generating domestic pressure on other states.[4]

Institutions prove to be one of the means which the emerging, regional & small powers use for power transitions and acclimate them for own political and economic benefits. Multi-faceted and multilateral nature of international institutions like IMF (International Monetary Fund) & UN help emerging powers to drive an agenda and make the weak member states react to it by taking advantage of institutional norms and rules. Nation states within themselves have asymmetric advantages of governance apparatus for mobilizing its population but lack collective

---

[1] Anders Wivel and T.V Paul "Exploring International Institutions and Power Politics" in Anders Wivel and T. V. Paul, eds., *International Institutions and Power Politics: Bridging the Divide* (Washington, DC: Georgetown University Press, 2019) pp. 2-19
[2] Ibid.
[3] Ibid.
[4] Norrin M. Ripsman " A Neoclassical Realist Explanation of International Institutions" in Anders Wivel and T. V. Paul, eds., *International Institutions and Power Politics: Bridging the Divide* (Washington, DC: Georgetown University Press, 2019) pp. 41-52

mobilization across sovereign borders.[5] International organizations provide them with a platform for collective mobilization to engage in a joint action.

3. **Strategies for institutional adjustment**. Rising powers, regional powers and small states employ certain strategies to force the institutions to adapt to the power shifts. As covered by Andreas Kruck and Bernhard Zangi, rising powers may resort to 'power bargaining' in which the they issue threats to the other states of the institution to accept compromise.[6] For example, during the financial crisis of 2012, China reached an agreement with the US on a more even-handed surveillance of the financial stability of IMF members (which was up till then directed at developing countries and exchange rate issues) after it threatened to disengage from IMF.[7] In some cases, 'rhetoric coercion' may be employed highlighting lack of legitimacy of institutions to make them accept adjustments.[8] In relation to China, amid the COVID-19 pandemic, Beijing provided a substantial illustration of abuse of the international system, compelling the World Health Organization to comply with Beijing's self-seeking preferences including the exclusion of Taiwan.[9] In some cases, emerging or regional powers engage in 'strategic co-optation' to mount a challenge thereby making material promises to buy defenders' agreement to institutional adjustments that fall in their common interests.[10] This was evident when China adopted an integrative strategy of persuasion and strategic co-optation to make its case for inclusion in

---

[5] Ripsman.
[6] Kruck, A. and Zangl, B. (2020), The Adjustment of International Institutions to Global Power Shifts: A Framework for Analysis. Glob Policy, 11: 5-16. https://doi.org/10.1111/1758-5899.12865
[7] Zangl, B., Heußner, F., Kruck, A. *et al.* Imperfect adaptation: how the WTO and the IMF adjust to shifting power distributions among their members. *Rev Int Organ* 11**,** 171–196 (2016). https://doi.org/10.1007/s11558-016-9246-z
[8] Kruck.
[9] *The Elements of the China Challenge by the Policy Planning Staff*, Office of the Secretary of State. (accessed November 12, 2021); available from https://www.state.gov/wp-content/uploads/2020/11/20-02832-Elements-of-China-Challenge-508.pdf
[10] Kruck.

Arctic Council as an observer status in 2013.[11] Emerging and established powers sometimes engage in 'principled persuasion' to challenge the institutional status quo and proclaim that institutional adjustments will lead to the improved legitimacy or efficiency of the institution to convince defenders that they have a joint interest in institutional adjustments.[12] CPEC (China Pakistan Economic Corridor) is an attempt by China to demonstrate itself as an attractive partner, prove its development model serves as a model to be exported and serve as an element of strategic competition with the US and India.[13]

4. **Processes of Power Politics and International Institutions.** In this section, I highlight various processes: soft balancing, information, wedging & networks and explore, how these processes affect are employed in the realm of power politics using international institutions.

4.1 **Soft Balancing.** Soft balancing is understood as attempts at restraining a threatening power through institutional delegitimization, as opposed to hard balancing, which relies on arms build-up and formal alignments.[14] Wivel and Paul argue that soft balancing through international institutions can be an effective means to peaceful change, spanning minimalist goals, aimed at incremental change without the use of military force and war, and maximalist goals, seeking transformation in the form of continuous interstate cooperation aimed at a more peaceful and just world order.[15] According to Paul, Soft balancing approaches include 'international institutions, economic statecraft, and diplomatic arrangement'.[16] In comparison to other soft-balancing strategies, "Institutional soft balancing" is typically more cost-effective and legitimate while allowing for a flexible response to aggression.[17] In the context of China, "Soft balancing" is undertaken by leveraging the use of SCO to influence an existing hegemonic power. China while not directly undertaking a challenge to US military power resort to cooperative military exercises within the SCO umbrella.[18] The joint SCO counter terrorism cooperative exercises demonstrate to the world that the region under China's lead has the capability & determination to fight regional terrorism and build as an alternate solution to countervailing the US military presence in the region. Soft power if seen from the perspective of Nye[19], is shaped by three sources: Culture, political values and foreign policies which all have been kept in mind while conceiving the BRI project. Developing states get attracted through the impact of BRI on the above mentioned sources and shape their preferences in relation to other available growth options.

---

[11] Stephen, M.D. and Stephen, K. (2020), The Integration of Emerging Powers into Club Institutions: China and the Arctic Council. Glob Policy, 11: 51-60. https://doi.org/10.1111/1758-5899.12834
[12] Kruck.
[13] Schwemlein, James. 2019. *Strategic implications of the China-Pakistan economic corridor*. https://purl.fdlp.gov/GPO/gpo149404.
[14] Wivel, A., & Paul, T. (2020). Soft Balancing, Institutions, and Peaceful Change. *Ethics & International Affairs*, 34(4), 473-485. doi:10.1017/S089267942000057X
[15] Ibid.
[16] Paul, T.V. (2005) Soft balancing in the age of US primacy. International Security 30 (1): 46–71. http://www.jstor.org/stable/4137458.

Another avenue in which China has been able to influence institutional framework is through the UN on the sustainability of the environment through the Belt and road initiative. The "Belt and Road Initiative Green Development Coalition" sets out a platform for increased integration of the countries who are part of the BRI and indirectly endorses & promotes the initiative. The coalition is an open, inclusive and voluntary international network which brings together the environmental expertise of all partners to ensure that the BRI brings long-term green and sustainable development to all concerned countries in support of the 2030 Agenda for Sustainable Development.[20] Its current set-up involves 134 partners including 26 Environmental Ministries of UN Member States and directly contributes to UN Environment's Medium-Term

---

[17] Wivel., 1.
[18] "*SCO "Peace Mission 2021"counter-terrorism drill concludes in Russia*" by *Global Times* accessed November 12, 2021; available from https://www.globaltimes.cn/page/202109/1235060.shtml
[19] Nye, Joseph S., Jr. (2004). *Soft power: The means to success in world politics*. New York: Public Affairs.
[20] "*The Belt and Road Initiative International Green Development Coalition*" (accessed November 13, 2021); available from https://www.unep.org/regions/asia-and-pacific/regional-initiatives/belt-and-road-initiative-international-green

Strategy and Programme of Work, cutting across several Sub-programmes.[21] A continued engagement with the UN helps China to exercise its soft power by promoting BRI under the guise of sustainable development goals to allure more states for joining the initiative and strengthen the voice of the initiative to its domestic audience.

4.2 **Information and leadership as means of power.**

Austin Carson and Alexander Thompson provides a detailed insight into how information plays an important role in enabling pursuit of power through international institutions depending on the information accessibility restrictions and information processing capabilities of different nations.[22]. They argue that information is an important source of power and that the design & practices of institutions have important consequences for translating material power disparities into policy.[23] China has taken information as the right source for changing the rhetoric of international institutions on the BRI and influencing states to change their inclination. An example of the same stems out from the misleading messages WHO was spreading during the COVID pandemic. Secretary General Tedros Adhanom Ghebreyesus, on many occasions, suggested that member states need not impose a travel ban from China, praised beijing's belated maladroit measures and understated the severity of the outbreak right up until it turned into a pandemic.[24]

Ripsman covers that leaders privilege their own interests and ideological goals by creating and joining organizations to demonstrate that the leadership is bringing prestige and status to their countries.[25] It is true for an authoritative regime like China in which its leader Xi Jin Ping appears to be using the BRI to stake his personalist claim to leadership. In line with Ripsman's claim, I would like to highlight the effect of leadership roles in UN agencies. Presently, four of the 15 UN specialized agencies are headed by Chinese nationals, which include" the United Nations Industrial Development Organization (UNIDP), the Food and Agriculture Organization (FAO), the International Civil Aviation Organization (ICAO) and the International Telecommunication Union (ITU). [26] An important thing to notice is that, since 2007, the position of under-secretary-general for the United Nations Department of Economic and Social Affairs (DESA) has been held by Chinese career diplomats (present one being Mr Liu Zhenmin since 2017)[27], giving the Chinese government opportunities to reshape the UN's development programs in accordance to its interests. By penetrating its leadership in the international organization like UN, China has been successful in influencing the opinion of these agencies in furtherance of its own ambitions. In order to garner strong positions in the international institutions, emerging powers may resort to exchange the leadership in international institutions by providing economic incentives. This was apparent, when prior to the election of the 9th Director-General of the Food and Agriculture Organization (FAO) in 2019, China slashed $78 million in debt owed by the Cameroon government whose nominated candidate coincidentally withdrew his candidature afterward. [28]

---

[21] Ibid.
[22] Austin Carson and Alexander Thompson " The Power in Opacity: Rethinking Information in International Organizations " in Anders Wivel and T. V. Paul, eds*., International Institutions and Power Politics: Bridging the Divide* (Washington, DC: Georgetown University Press, 2019) pp. 101-116
[23] Ibid.
[24] "How China Is Remaking the UN In Its Own Image", *The Diplomat* accessed November 13, 2021, https://thediplomat.com/2020/04/how-china-is-remaking-the-un-in-its-own-image/

---

[25] Norrin M. Ripsman "A Neoclassical Realist Explanation of International Institutions" in Anders Wivel and T. V. Paul, eds*., International Institutions and Power Politics: Bridging the Divide* (Washington, DC: Georgetown University Press, 2019) pp. 41-53
[26] "How China Is Remaking the UN In Its Own Image", *The Diplomat* accessed November 13, 2021,https://thediplomat.com/2020/04/how-china-is-remaking-the-un-in-its-own-image/
[27] Global Leadership Team, Secretariat and Regional commissions (UN) (accessed November 13, 2021); available from https://www.un.org/sg/en/global-leadership/department-of-economic-and-social-affairs/all
[28] Ibid.

## 4.3 **Symbolic capital and Networks as sources of Power**

As covered by Daniel H. Nexon, position in the hierarchy of a field (Spheres of social Action) comes from the accumulation of field relevant capital i.e assets and performances which comes in three forms of economic, social and resources generated by network ties.[29] Great power have the ability to convert economic and military capital into diplomatic capital and set the rate of exchange within and across the fields. With the second largest economy and the largest military in the world alongwith membership in international institutions as a form of symbolic capital, China possesses the capability to shape global economy. Beijing considers AIIB (Asian Infrastructure Investment Bank) and the Silk road funds as part of the production capacity cooperation to reinvigorate the global economic growth. It serves as a symbolic capital for China to demonstrate its status of a great power and a future hegemon to consolidate its standing amongst international & domestic audiences.

Stacie Goddard contend that institutional affiliation in terms of social ties & networks positions influence the nature of states and serve as toolkits available to revisionist states. [30] Discussing how revisionist politics play out in different institutional orders, she argues that revisionism is not limited to aggressive attacks on the institutional order and one can nuance our understanding of the relationship between revisionist policies and institutional change by drawing institutions as networks.[31] Through the BRI and other associated networks within the framework of regional diplomacy, China attracts and exposes other countries to economic cooperation through the lens of common prosperity. It is evident that China is also seeking to build its own network of bases by strategic partnership with the countries joining BRI to carry out dominance. Its approach lays the ground work for military utilization of the facilities like "port park city" development model that integrates the port with industrial parks and support industries like shipbuilding and resupply services that enhance the port's capacity to support PLA naval ships.[32]

---

[29] Daniel H. Nexon " International order and power politics" in Anders Wivel and T. V. Paul, eds., *International Institutions and Power Politics: Bridging the Divide* (Washington, DC: Georgetown University Press, 2019) pp. 197-214
[30] Stacie Goddard " Revisionists, Networks and the Liberal Institutional order" in Anders Wivel and T. V. Paul, eds., *International Institutions and Power Politics: Bridging the Divide* (Washington, DC: Georgetown University Press, 2019) pp. 117-136
[31] Ibid.

4.4 **<u>Wedging : Mechanism to Prevent Balancing Coalitions.</u>** Timothy Crawford defines wedging as a "state's attempt to prevent, break up, or weaken a threatening or blocking alliance at an acceptable cost."[33] When it is successful, the state (the divider) gains advantage by reducing the strength of enemies organized against it, can turn opponents into neutrals or allies to defend own alliances and thereby trigger surprising power shifts with significant consequences for war and peace and the trajectory of international politics.[34] By feeding increased BRI investments in countries in the Indo pacific, Beijing uses economic instruments to prevent a balancing coalition from emerging as BRI is a major source of investments which traps developing states into large debt repaying thereby making them ineffective and unlikely to joining other alliances. The smaller states of South Asia − Bangladesh, Sri Lanka, Nepal and Maldives − have been effectively wedged from other regional powers like India or coalitions like QUAD, by the BRI & other economic/infrastructure aid that China offers.[35]

---

[32] "*Weaponizing the Belt and Road initiative*" by Asia Society Policy Institute (accessed November 14, 2021); available from https://asiasociety.org/policy-institute/weaponizing-belt-and-road-initiative
[33] Crawford, Timothy W. "Preventing Enemy Coalitions: How Wedge Strategies Shape Power Politics." *International Security* 35, no. 4 (2011): 155–89. http://www.jstor.org/stable/41289683.
[34] Ibid.
[35] Paul, T. V. (2019). Why Balancing Towards China is not Effective : Understanding BRI's Strategic Role. (RSIS Commentaries, No. 049). RSIS Commentaries. Singapore: Nanyang Technological University.

5. **Conclusion.** During the COVID 19 pandemic, China has been continuously engaging to strengthen its position within the system of international and regional institutions to garner support for its ultimate goal to emerge as a global power and weaken the global position of the US. The multilateral nature of the international institutions provide a platform to engage with developing countries to intensify its economic cooperation especially through BRI.

With large leaps into UN peacekeeping operations and international institutions like ISO (International Standards Organisation), ITU (International Telecommunication Union) and IEC (International Electrotechnical Commission), China's aggressive attempts to make international institutions as a launch pad in achieving the great power status would have a negative impact and will render the stature of these institutions as ineffective. Countries with strong governments need not come under the pressure of being part of the BRI as it will increase the flow of foreign labour and increase the financial burden. BRI if seen from a vantage point of interests, promotes the national interests of China more rather than affected countries. In the coming years, BRI will remain priority for Beijing but the geopolitical risks are likely to persist. As China continues to recover from the impact of the pandemic, its reaction to the norms of the international institutions can have implications in the present geopolitics of the world and in particular to US led liberal order. It is likely that Beijing will consolidate its position & gains to accelerate BRI.

As evident from our analysis, in future, various types of states (e.g., great powers, regional powers, or smaller states) will continue to utilize international institutions in the pursuit of power politics till the time it remains a cost effective solution for balancing and exerting their influence. A continued creative moves in power politics to prevent institutional decay and foster institutional renewal will require efforts from states for a better peaceful transition.